\RequirePackage{fix-cm}
\documentclass[smallextended]{svjour3}       
\smartqed  
\usepackage{graphicx}
\usepackage{amssymb}
\begin{document}

	\newcommand {\nc} {\newcommand}
	\nc {\beq} {\begin{eqnarray}}
	\nc {\eeq} {\nonumber \end{eqnarray}}
	\nc {\eeqn}[1] {\label {#1} \end{eqnarray}}
\nc {\eol} {\nonumber \\}
\nc {\eoln}[1] {\label {#1} \\}
\nc {\ve} [1] {\mbox{\boldmath $#1$}}
\nc {\ves} [1] {\mbox{\boldmath ${\scriptstyle #1}$}}
\nc {\mrm} [1] {\mathrm{#1}}
\nc {\half} {\mbox{$\frac{1}{2}$}}
\nc {\thal} {\mbox{$\frac{3}{2}$}}
\nc {\fial} {\mbox{$\frac{5}{2}$}}
\nc {\la} {\mbox{$\langle$}}
\nc {\ra} {\mbox{$\rangle$}}
\nc {\etal} {{\emph{et al.}}}
\nc {\eq} [1] {(\ref{#1})}
\nc {\Eq} [1] {Eq.~(\ref{#1})}
\nc {\Sec} [1] {Sec.~\ref{#1}}
\nc {\tbl} [1] {Table~\ref{#1}}
\nc {\Fig} [1] {Fig.~\ref{#1}}
\nc {\Sch} {Schr\"odinger }
\nc {\flim} [2] {\mathop{\longrightarrow}\limits_{{#1}\rightarrow{#2}}}


\title{Combining Halo-EFT descriptions of nuclei and precise models of nuclear reactions
}
\subtitle{Celebrating 30 years of Steven Weinberg's papers on Nuclear Forces from Chiral
Lagrangians}


\author{Pierre Capel         
}


\institute{P.\ Capel \at
              Institut f\"ur Kernphysik, Johannes Gutenberg-Universit\"at Mainz, Johann-Joachim-Becher Weg 45, 55099 Mainz, Germany\\
              Tel.: +49 (0)6 131 39 29628\\
              \email{pcapel@uni-mainz.de}           
}

\date{Received: date / Accepted: date}

\maketitle

\begin{abstract}
The clear separation of scales observed in halo nuclei between the extended halo and the compact core makes these exotic nuclei a perfect subject for Effective Field Theory (EFT).
Such description leads to a systematic expansion of the core-halo Hamiltonian, which naturally orders the nuclear-structure observables.
In this short review, I show the advantage there is to include Halo-EFT descriptions within precise models of reactions.
It helps identifying the nuclear-structure observables that matter in the description of the reactions, and enables us to easily bridge predictions of nuclear-structure calculations to reaction observables.
I illustrate this on breakup, transfer and knockout reactions with $^{11}$Be, the archetypical one-neutron halo nucleus.
\end{abstract}

\section{Introduction}\label{Intro}
The seminal development of Steven Weinberg in the early 1990s has significantly modified the way people approach the notion of nuclear force and their use in modern nuclear physics \cite{Wei90,Wei91}. 
His idea of Effective Field Theory (EFT) is rooted in the vision that short-range physics does not matter at first orders.
One rather unexpected application of EFT is the halo structure in nuclear physics \cite{BHvK02,BHvK03}; see Ref.~\cite{HJP17} for a recent review on Halo-EFT.

Discovered in the mid 1980s \cite{Tan85b,Tan85l}, halo nuclei are characterised by a matter radius significantly larger than their isobars' \cite{Tan96}.
This unusual size results from a strong cluster structure.
Thanks to their loose binding to the nucleus, one or two valence neutrons exhibit a high probability of presence at a large distance from the other nucleons, and form a sort of extended and diffuse \emph{halo} around the tight and compact core of the nucleus \cite{HJ87}.
Examples include $^{11}$Be and $^{15}$C with one neutron in their halo, and $^{11}$Li, a two-neutron halo nucleus.
This clear separation of scales makes halo nuclei ideal candidates for an EFT \cite{BHvK02,BHvK03,HJP17}.

The degrees of freedom of Halo-EFT are the halo neutron(s) and the core, whose internal structure is neglected.
Following the EFT idea, the core-halo Hamiltonian is expanded in powers of the small parameter, which corresponds to the ratio of the small size of the core divided by the large radius of the halo.
The idea behind that expansion is that the short-range physics does not matter in most cases, and that it can be absorbed in the Low-Energy Constants (LECs) of the theory, viz. the coefficients of that expansion.
The LECs are fitted to known low-energy properties of the halo structure, such as its binding energy and the Asymptotic Normalisation Constant (ANC) of the wave function of the core-halo bound state.
These constraints can be derived from experiment or from precise nuclear-structure calculations.
The archetypical one-neutron halo nucleus, $^{11}$Be, for example, has recently been computed \emph{ab initio} by Calci \etal\ within the No-Core Shell Model with Continuum (NCSMC) \cite{CNR16}.
Accordingly, we can adjust the Halo-EFT description of that nucleus with a highly educated guess of its nuclear-structure observables.

Because of their short lifetime, these exotic nuclei are studied mostly through reactions.
In breakup, the loose binding between the halo neutron(s) and the core is broken through the interaction with a target \cite{Fuk04,Pal03,BC12}, hence revealing the internal structure of the projectile.
In $(\rm d,p)$ transfer on the core of the nucleus, $c$-n halo states can be populated \cite{Sch12,Sch13}.
The analysis of these reactions provides useful nuclear-structure information \cite{GCM14}.
Finally, knockout has been extensively used to study halo structures \cite{Aum00,HT03}.
There, the halo neutron is removed at high beam energy on a light target, like $^{9}$Be.
The shape of the momentum distribution of the core stripped from its halo neutron(s) provides valuable information on the halo wave function \cite{Aum00,HT03}.

Recently, efforts to describe both the halo nucleus and the reaction process within EFT have been made \cite{HP11,AP13,SPH19}.
Although the perturbative approach used in these papers to simulate the reaction provides a correct qualitative estimate of the process, higher-order effects, such as couplings within the continuum \cite{EB96,EBS05,CB05} and Coulomb-nuclear interferences \cite{HLN06}, need to be accounted for to reach the quantitative level.
The studies I summarise in this review aim at including a Halo-EFT description of the nucleus within precise models of reactions, which account for all orders of the process \cite{CPH18,YC18,MC19,MYC19,HC21}.

A key question in the analysis of experiments with halo nuclei is to know which structure observables are actually probed during the reaction.
Various studies using precise models of reaction have already been performed \cite{CN06,CN07}.
However, thanks to its systematic expansion of the core-halo Hamiltonian, Halo-EFT provides us with a very natural tool to explore this question.
By looking at the changes induced by the various terms in the EFT expansion on the reaction observables, we can study the influence of the projectile description upon the reaction mechanism.
In addition, because the LECs can be constrained from accurate nuclear-structure calculations, this idea can also help us bridge these theoretical results and reaction measurements.
In this review, I illustrate on $^{11}$Be how this can be done for breakup in \Sec{BU}, transfer in \Sec{Trans} and knockout in \Sec{KO}. I offer my conclusions in \Sec{Conclusion}.

\section{Description of one-neutron halo nuclei within Halo-EFT}\label{HaloEFT}

As mentioned in the Introduction, halo nuclei exhibit a strongly clusterised structure, in which one---or two---neutron decouples from the core of the nucleus.
Such a structure is well accounted for within a few-body model of the nucleus: a core $c$, whose internal structure is ignored, to which a neutron n is loosely bound.
The Hamiltonian that describes such a two-body model of one-neutron halo nuclei reads
\beq
H_0=-\frac{\hbar^2}{2\mu}\Delta+V_{c\rm n}(r),
\eeqn{e1}
where $\ve{r}$ is the coordinate of the halo neutron relative to the core, $\Delta$ is the corresponding Laplacian, $\mu$ is the $c$-n reduced mass and $V_{c\rm n}$ is an effective potential that simulates the $c$-n interaction.
The eigenstates of $H_0$ describe the $c$-n structure of the halo nucleus.
The negative-energy states are discrete and correspond to $c$-n bound states, whereas the positive-energy eigenstates of $H_0$ form the continuum of the halo nucleus, i.e., they describe the states in which the neutron is dissociated from the core.

\begin{figure}
  \includegraphics[width=0.5\textwidth]{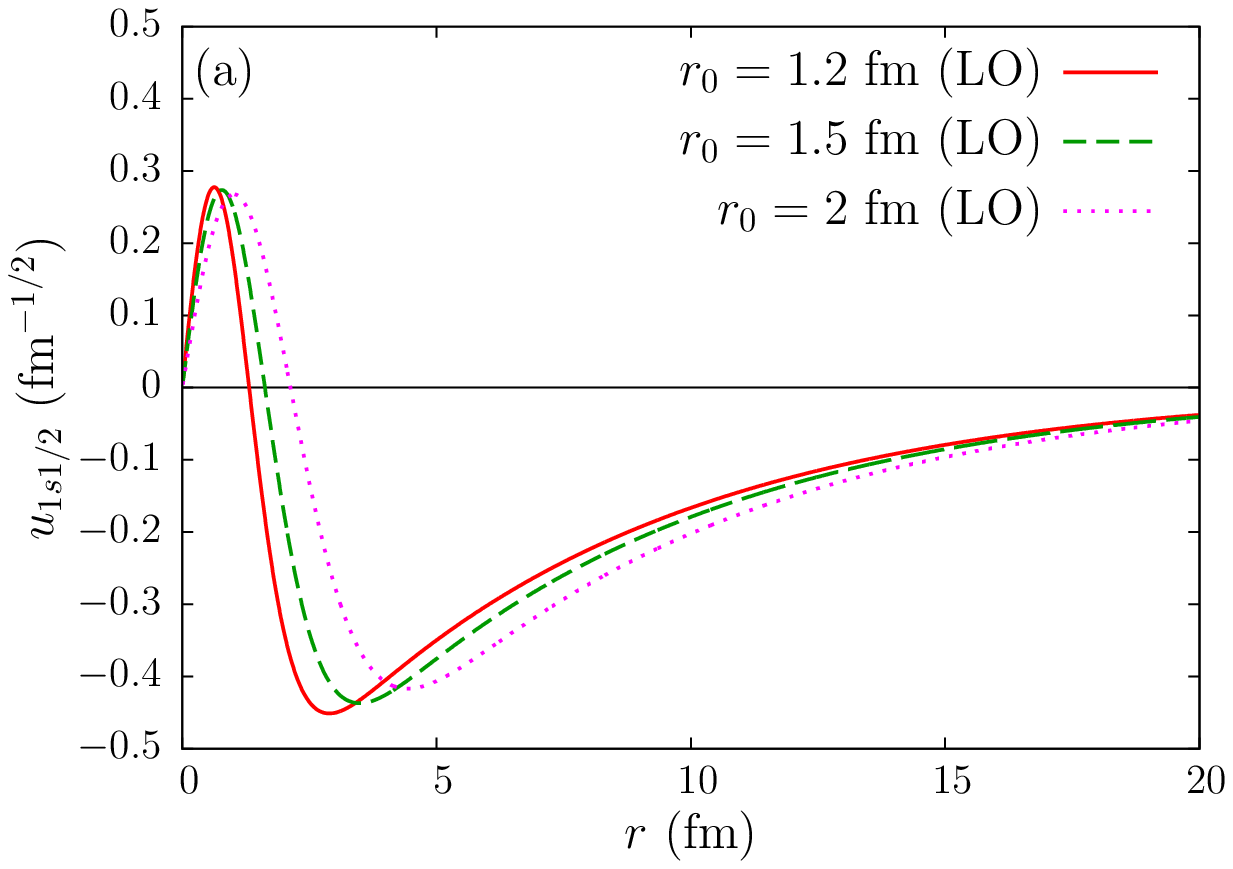}
  \includegraphics[width=0.5\textwidth]{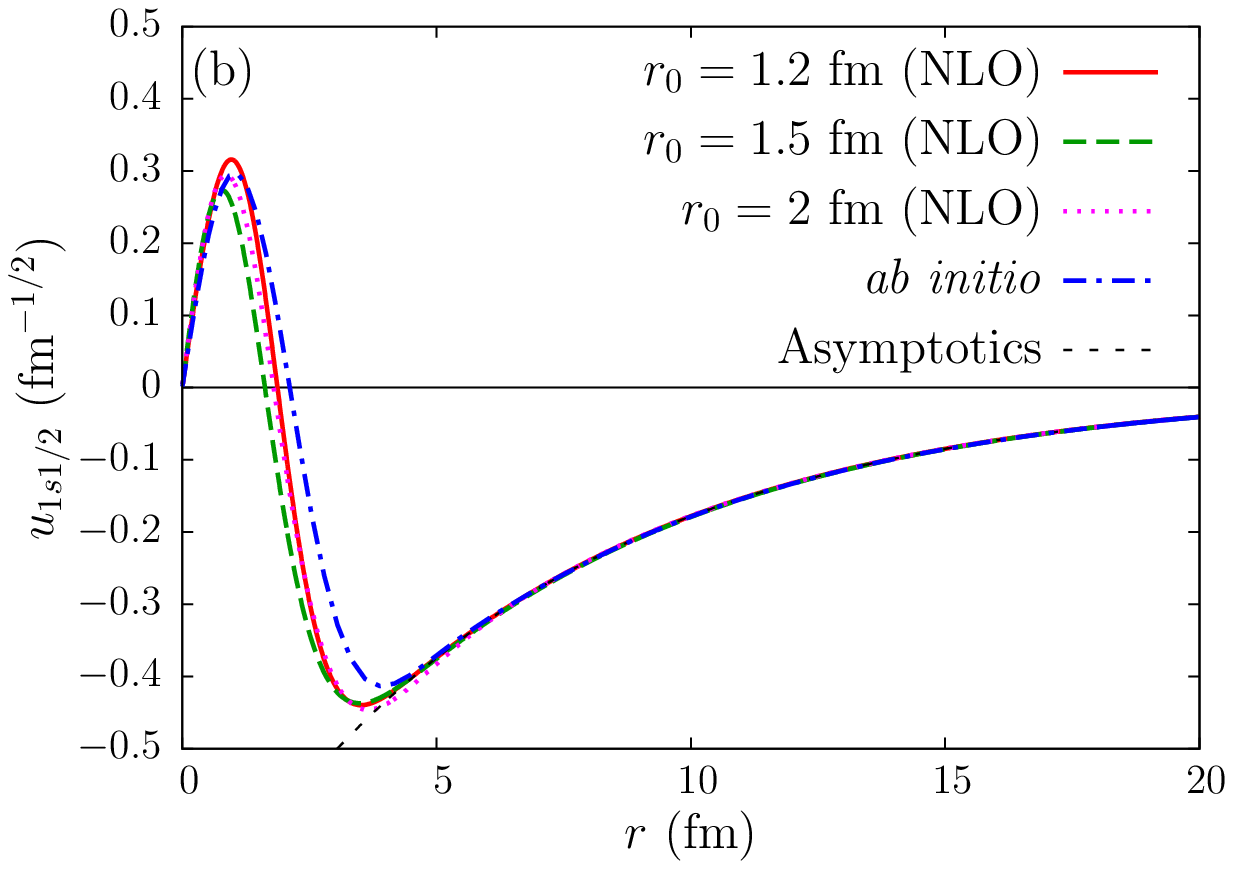}
\caption{Reduced radial wave functions for the $\half^+$ ground state of $^{11}$Be described within Halo-EFT as a $^{10}$Be core in its $0^+$ ground state to which a neutron is loosely bound in a $1s_{1/2}$ orbital. (a) At LO, where only one LEC is adjusted to reproduce the experimental $S_{\rm n}$; results for three values of the cutoff $r_0$ are shown. (b) At NLO, where two LECs are fitted to reproduce the experimental $S_{\rm n}$ and the ANC of the \emph{ab initio} overlap wave function.
Reprinted figure with permission from Ref.~\cite{CPH18} Copyright (2018) by the American Physical Society.
}
\label{f1}
\end{figure}

In most models of reactions, $V_{c\rm n}$ is chosen of Woods-Saxon form, whose parameters are fitted to reproduce the low-energy spectrum of the nucleus \cite{BC12}, i.e., the energy, spin and parity of its first states.
For the archetypical one-neutron halo nucleus $^{11}$Be, the $\half^+$ ground state is described by a $1s_{1/2}$ neutron bound to a $^{10}$Be core in its $0^+$ ground state by a mere 503~keV.
This choice of a $^{10}$Be-n ground-state wave function which exhibits a node ($n_r=1$) is common in nuclear-cluster physics.
It accounts for the presence of neutrons inside the core, which occupy the $0s_{1/2}$ orbital in a mean-field description of the nucleus.
In a microscopic model, such as the \emph{ab initio} NCSMC of Calci \etal\ \cite{CNR16}, a node appears in the ground-state overlap wave function as a result of Pauli's antisymmetrisation principle, see the blue dash-dotted line in \Fig{f1}(b).
As was shown in Ref.~\cite{CBM03b}, the presence of Pauli forbidden states has no effect on breakup calculations.
The results presented below would thus not change had we used nodeless wave functions, more usual in EFT approaches.
Above its ground state, the $^{11}$Be $\half^-$ first and only one excited bound state is reproduced as a $0p_{1/2}$ neutron  bound to $^{10}$Be$(0^+)$ by 184~keV.
Note that this order is inverted compared to an extreme shell-model vision of the nuclei.

Above the one-neutron threshold, the $\fial^+$ state at $E=1.274$~MeV is usually described as a $d_{5/2}$ single-particle resonance \cite{CGB04}.
Further up in the continuum, the $\thal^+$ state at $E=2.90$~MeV can be tentatively described as a $d_{3/2}$ resonance.
However, the vicinity of that state to the threshold for the excitation of the $^{10}$Be core to its first $2^+$ excited state, makes that approximation less reliable.

The \emph{ab initio} calculation of Calci \etal\ has confirmed this single-particle description of $^{11}$Be \cite{CNR16}.
It also provides theoretical predictions for the ANC of both bound states and the phaseshift in the low-energy continuum, which, as we will see below, is a key input for the description of reactions with $^{11}$Be.

Because EFT is designed to be insensitive to the short-distance details, each term of its expansion is taken to be a contact term, or its derivatives.
The LECs, i.e., the magnitude of each of these terms, are fitted to reproduce known structure observables.
They can be experimental data, such as the one-neutron binding energy, or theoretical predictions, such as the ANC of the bound states.
The breakdown scale of this EFT is the size of the core.

In practice, the contact interactions used to expand $V_{c \rm n}$ are regularised by a Gaussian of range $r_0$, which corresponds to the scale of the short-range physics neglected in the description of the nucleus.
If the Halo-EFT idea is correct, the results of the reaction calculations should be independent of the exact value of $r_0$, as long as it remains small.
The $c$-n interaction is parametrised partial wave per partial wave.
At Leading Order (LO), it contains only one term within the sole $s$ wave \cite{HJP17}:
\beq
V_{c \rm n}^{\rm LO}(r)=V_{s1/2}^{(0)}\ e^{-\frac{r^2}{2r_0^2}}.
\eeqn{e2}
The depth $V_{s1/2}^{(0)}$ is then fitted to reproduce the binding energy of the halo neutron to the nucleus $S_{\rm n}$, in the case of $^{11}$Be that energy is 0.503~MeV.
This fixes the exponential decay of the asymptotics of the $^{11}$Be ground state radial wave function, as can be seen in \Fig{f1}(a), which depicts the reduced radial wave function of the $1s_{1/2}$ single-particle state modelling $^{11}$Be $\half^+$ ground state at LO \cite{CPH18}.
Three different values of the cutoff are shown: $r_0=1.2$~fm (solid red line), 1.5~fm (green dashed line) and 2~fm (magenta dotted line).
These wave functions differ significantly in their internal part, i.e., for $r\lesssim5$~fm.
Note also that although they exhibit the same exponential decay, the normalisation of that tail---the ANC---varies with $r_0$.

At Next-to-Leading Order (NLO), the development extends to both $s$ and $p$ waves, and exhibits two terms: a Gaussian and its second-order derivative \cite{HJP17}.
For practical use, the following expansion is considered \cite{CPH18}:
\beq
V_{c \rm n}^{\rm NLO}=V_{lj}^{(0)}\ e^{-\frac{r^2}{2r_0^2}}+V_{lj}^{(2)}\ r^2\,e^{-\frac{r^2}{2r_0^2}}.
\eeqn{e3}
To adjust the depths $V_{lj}^{(0)}$ and $V_{lj}^{(2)}$ in the $s_{1/2}$ partial wave, we consider, in addition to the halo-neutron separation energy, the ANC predicted by the \emph{ab initio} calculation of Calci \etal\ \cite{CNR16}.
This fixes not only the decay constant of the asymptotics of the wave function, but also its normalisation.
This is illustrated in \Fig{f1}(b), where, in addition to the Halo-EFT radial wave functions at NLO for the three values of $r_0$, the overlap wave function obtained in the \emph{ab initio} calculation is shown in blue dash-dotted line \cite{CNR16}.
In that case, the differences between the single-particle wave functions are significantly reduced.
Nevertheless, they remain noticeable in the interior of the nucleus.

At NLO, $V_{c\rm n}$ has also to be constrained in the $p$ waves.
In $p_{1/2}$, the two LECs can be adjusted to reproduce the binding energy and ANC of the $\half^-$ excited bound state of $^{11}$Be.
The binding energy is well known experimentally, and that state ANC is taken once again from Calci \etal 's calculation \cite{CNR16}.
In the $p_{3/2}$ partial wave, there is no low-energy state upon which the parameters of the interaction \eq{e3} can be fitted.
Therefore, we rely solely on the \emph{ab initio} calculation, which predicts a $\thal^-$ phaseshift compatible with 0 at low energy \cite{CNR16}.
Accordingly, we set $V_{c\rm n}=0$ in the $p_{3/2}$ partial wave.


\section{Breakup}\label{BU}

The breakup of $^{11}$Be, i.e., its dissociation into $^{10}$Be and a neutron, has been measured on Pb and C at about $70A$~MeV at RIKEN \cite{Fuk04} and at $520A$~MeV at the GSI \cite{Pal03}.
To study these reactions theoretically, we consider the usual few-body model of reactions \cite{BC12}.
The projectile $P$ is seen as a two-body system, whose structure is described by Halo-EFT as explained in \Sec{HaloEFT}.
The structure of the target $T$ is ignored and its interaction with the projectile constituents (the core $c$ and the halo neutron n) is simulated by optical potentials $V_{cT}$ and $V_{{\rm n}T}$, which describe, respectively, the $c$-$T$ and n-$T$ elastic scatterings.

Within this description of the reaction, studying the $P$-$T$ collision reduces to solving the corresponding three-body \Sch equation with the condition that the projectile, initially in its ground state, is impinging on the target.
To solve that problem at the RIKEN energy, we use the Dynamical Eikonal Approximation (DEA) \cite{BCG05,GBC06}, which is based on the eikonal approximation \cite{Glauber}, but does not rely on the adiabatic---or sudden---approxima\-tion considered in the usual eikonal model of collisions.
This enables us to properly account for the dynamics of the reaction \cite{CEN12}, including couplings within the continuum and Coulomb-nuclear interferences mentioned in the Introduction, which lack in perturbative approaches of breakup \cite{HP11,AP13}.
As shown in Refs.~\cite{BCG05,GBC06}, this model of breakup provides excellent agreement with various experimental data at intermediate beam energies, viz. around $70A$~MeV.

\begin{figure}
  \includegraphics[width=0.5\textwidth]{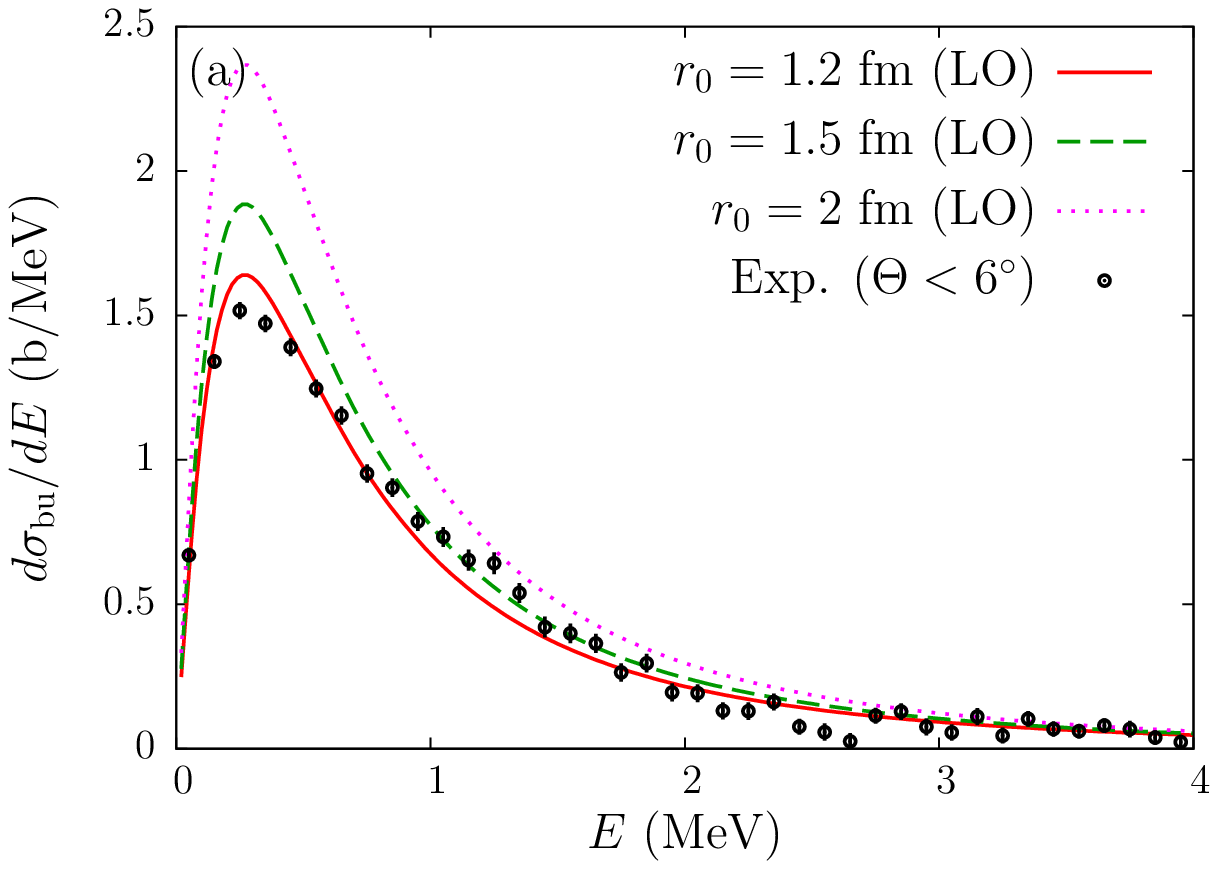}
  \includegraphics[width=0.5\textwidth]{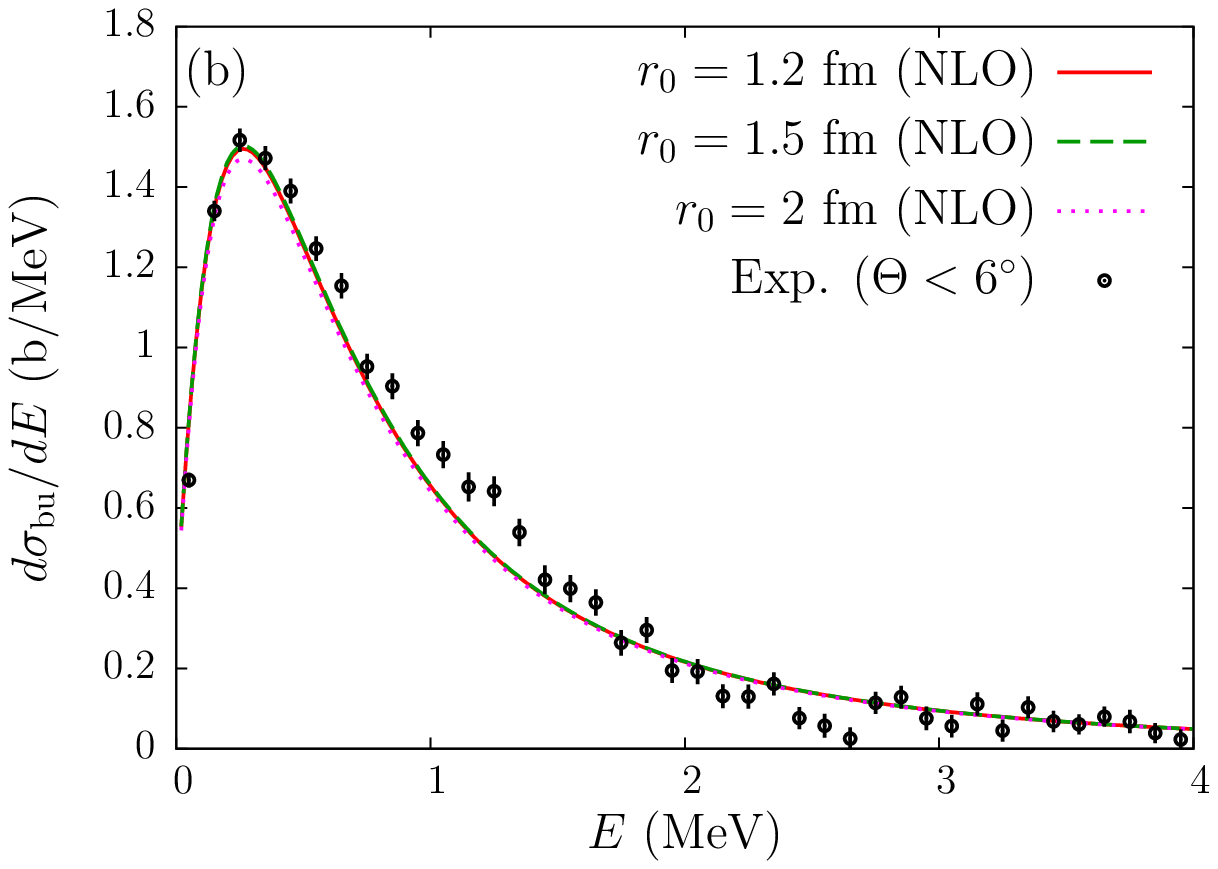}
\caption{Coulomb breakup of $^{11}$Be on Pb at $69A$~MeV plotted as a function of the relative energy $E$ between the $^{10}$Be and the halo neutron after dissociation.
The calculations are performed within the DEA with different ranges $r_0$ for the $^{10}$Be-n Halo-EFT interaction.
(a) Using a LO Halo-EFT description of $^{11}$Be.
(b) Using a $^{10}$Be-n potential at NLO \cite{CPH18}.
Data are from Ref.~\cite{Fuk04}.
Reprinted figures with permission from Ref.~\cite{CPH18} Copyright (2018) by the American Physical Society.
}
\label{f2}
\end{figure}

The breakup cross section measured at RIKEN on lead at $69A$~MeV is shown in \Fig{f2} as a function of the relative energy $E$ between the $^{10}$Be core and the halo neutron after dissociation \cite{Fuk04}.
In addition to the data, panel (a) shows the DEA cross sections obtained with LO Halo-EFT descriptions of $^{11}$Be.
The three cutoffs shown in \Fig{f1}(a) are considered here with the same line types.
All calculations reproduce the general energy dependence of the data: they peak at $E\approx0.3$~MeV before quickly decaying at larger energies.
This rough agreement with experiment can be qualitatively explained at the first order of the perturbation theory.
Although higher-order effects, such as couplings within the continuum and Coulomb-nuclear interferences, are present in Coulomb breakup \cite{EB96,EBS05,CB05,HLN06}, this reaction is dominated by a one step E1 transition from the initial $s$ ground state to the $p$ continuum.
For such a transition, the general shape of the energy distribution, in particular the position of its maximum and its magnitude, is mostly constrained by the binding energy of the nucleus \cite{TB04,TB05}.
The DEA calculations confirm that including a realistic estimate of the nuclear interaction between the projectile constituents and the target and properly accounting for the projectile dynamics within the reaction model do not change qualitatively the conclusions of Refs.~\cite{TB04,TB05}.

On a quantitative level, this sole input in the projectile description is not sufficient, as illustrated in \Fig{f2}(b).
Once the Halo-EFT description of $^{11}$Be is constrained at NLO, i.e., once it is fitted to the ANC of the initial bound state and the phaseshift in the $^{10}$Be-n $p_{3/2}$ and $p_{1/2}$ continua, the DEA calculations no longer depend on the cutoff $r_0$ \cite{CPH18}.
Moreover they are in perfect agreement with the RIKEN data \cite{Fuk04}.
This indicates that this reaction observable is not significantly sensitive to details beyond NLO in the description of $^{11}$Be.
This has been confirmed in Ref.~\cite{CPH18}, where it is shown that adding details beyond NLO, such as $d$ resonances, does not modify the results shown in \Fig{f2}(b).
Moreover, because this reaction is Coulomb dominated, the uncertainty of the calculations due to the particular choice of the optical potentials $V_{cT}$ and $V_{{\rm n}T}$ remains very small \cite{CBM03c}.

Due to the excellent agreement with the data, and since the DEA includes higher-order effects and nuclear $P$-$T$ interactions, 
this result also validates the \emph{ab initio} predictions used to fit the LECs of the model, i.e., the bound-state ANCs and the phase-shift in the $p$ waves.
Note that, as found in Ref.~\cite{CN06}, the phaseshift in both $p$ waves influences the result; the sole value of the ANC for the $1s_{1/2}$ bound state is not sufficient to fix the magnitude of the breakup cross section \cite{CPH18}.
Therefore it is necessary to consider the NLO description of $^{11}$Be to constrain $V_{c\rm n}$ not only in the $s$ wave, but also in the $p$ wave to reproduce the correct $c$-n continuum.
As mentioned earlier, the phaseshift in the $d$ wave has little influence on the result, indicating that the NLO description of $^{11}$Be is also sufficient to reproduce this reaction observable.
This enables us to conclude, as shown earlier \cite{CN07}, that the Coulomb breakup of one-neutron halo nuclei is not sensitive to the short-range physics of the projectile.
It is therefore illusionary to attempt to infer from such measurements any information depending on the internal part of the overlap wave function, such as the spectroscopic factor.

\begin{figure}
  \includegraphics[width=0.5\textwidth]{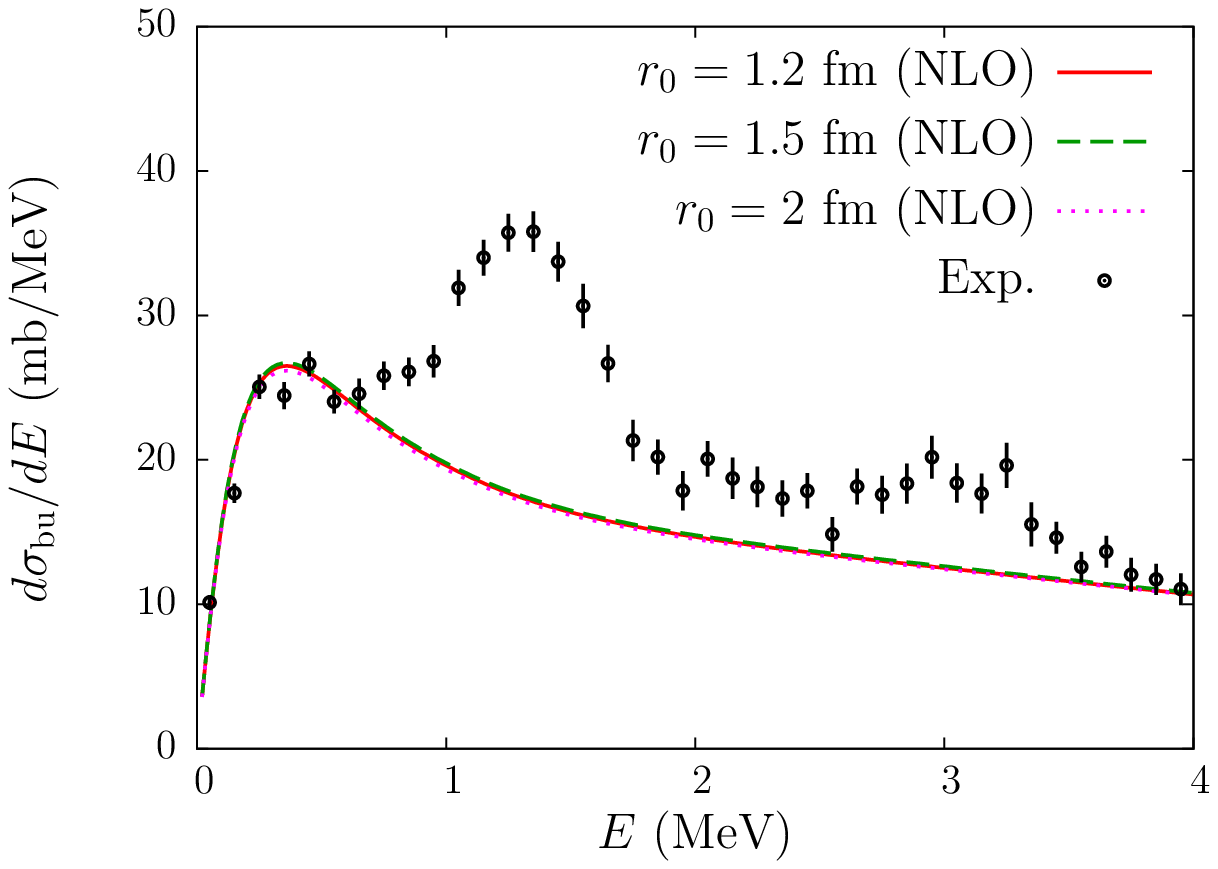}
  \includegraphics[width=0.5\textwidth]{ReviewFBf3b.eps}
\caption{
Left panel: breakup cross section of $^{11}$Be into $^{10}$Be and a neutron on C at $67A$~MeV plotted as a function of the relative energy $E$ between the two fragments.
Calculations are performed within the DEA with NLO $^{10}$Be-n effective potentials.
Data are from Ref.~\cite{Fuk04}.
Reprinted figures with permission from Ref.~\cite{CPH18} Copyright (2018) by the American Physical Society.
Right panel: energy distribution for the breakup of $^{11}$Be on Pb (top) and C (bottom) at $520A$~MeV.
CCE calculations with a relativistic treatment of the $P$-$T$ motion \cite{MC19} are compared to the GSI data \cite{Pal03} using the same NLO descriptions of $^{11}$Be. 
Reprinted figures from Ref.~\cite{MC19} with permission from Elsevier.
}
\label{f3}
\end{figure}

In addition to the Coulomb breakup of $^{11}$Be, Fukuda \etal\ have also measured its nuclear breakup on carbon at $67A$~MeV \cite{Fuk04}.
The energy distribution for that reaction is displayed in the left panel of \Fig{f3}.
Besides the experimental data, that plot includes the results of our DEA calculations using the NLO descriptions of $^{11}$Be already considered on Pb \cite{CPH18}. 
Once again, we see that all cutoffs $r_0$ lead to nearly identical cross sections, confirming the previous result of Ref.~\cite{CN07} that these reactions are purely peripheral, in the sense that the probe only the tail of the projectile.

Compared to the Coulomb-dominated case, the agreement with the data is less good.
Although our dynamical calculations reproduce the correct magnitude of the cross section and its general shape, i.e., its low-energy maximum and its slow decay at larger energy, they do not reproduce the two peaks at about 1.3~MeV and 3~MeV in the $^{10}$Be-n continuum.
These peaks correspond to the $\fial^+$ and $\thal^+$ states mentioned earlier \cite{Fuk04,CGB04}.
Adding them as single-particle $d_{5/2}$ and $d_{3/2}$ resonances within a beyond-NLO description of $^{11}$Be slightly improves the situation in the region of the first resonance, but it barely makes any difference in the second's \cite{CPH18}.
Some degrees of freedom, not included in our model of $^{11}$Be, must be significant in the breakup towards these states.
The first of such degrees of freedom is the $^{10}$Be core structure, and in particular the presence of a $2^+$ excited state at 3.37~MeV above its $0^+$ ground state.
The lack of breakup strength in the vicinity of the resonances is therefore most likely due to the dynamical excitation of the core during the reaction.
This idea has been successfully explored by Moro and Lay in Ref.~\cite{ML12} within a DWBA reaction model that explicitly includes core excitation in the description of the projectile.
These authors have shown that the core excitation to its first $2^+$ excited state plays a key role in the population of these two resonances during the reaction.
We have recently confirmed that possibility by simulating the virtual excitation of $^{10}$Be during the reaction through the use of a three-body force \cite{CPH20a}.
This development follows the original idea of Fujita and Miyazawa to account for the excitation of the $\Delta(1232)$ resonance in nucleon-nucleon collisions \cite{FM57}.
The good agreement with the data obtained in this way confirms the result of Moro and Lay, while illustrating the interest of EFT approaches in the description of nuclear reactions.

To be complete, it is worth noting the significant influence the optical potentials $V_{cT}$ and $V_{{\rm n}T}$ have on the theoretical cross sections for the breakup on such a light target, viz. when the process is nuclear dominated.
As shown in Ref.~\cite{CGB04}, the cross section can vary up to a factor of two when $V_{cT}$ is changed.
However, the potential considered here has been adjusted to reproduce the elastic scattering of $^{10}$Be on $^{12}$C at $59.4A$~MeV \cite{ATB97}, which is very close to the conditions of the RIKEN experiment.
Accordingly we can assume that it is reliable enough not to lead to the large uncertainty observed in Ref.~\cite{CGB04}.

The breakup of $^{11}$Be on both Pb and C has also been measured at the GSI, although at the much higher beam energy of $520A$~MeV  \cite{Pal03}.
The right side of \Fig{f3} shows the energy distributions inferred from that experiment on Pb (upper panel) and C (lower panel).
Following the excellent results obtained at the lower RIKEN energy, we have considered the NLO descriptions of $^{11}$Be presented earlier to see if they would also reproduce these high-energy measurements.
To describe the reaction, we have used the Coulomb-Corrected Eikonal model (CCE) \cite{CBS08} with a relativistic treatment of the $P$-$T$ motion \cite{MC19}.
Here also, all three theoretical cross sections are nearly superimposed to one another, confirming the negligible dependence of these reaction observables to the short-range physics of the projectile structure.

The CCE cross sections obtained with these NLO inputs are in good agreement with the data.
On Pb they slightly overestimate the data at the low-energy peak, but they reproduce nicely the high-energy decay observed experimentally.
On C, we note once more the lack of breakup strength in the region of $\fial^+$ resonance, since it is not included on our NLO model space.

Coupling a Halo-EFT description of $^{11}$Be to precise models of breakup is interesting on two fronts.
First it enables us to determine which structure observables are probed by the reaction.
On Pb, a NLO description is necessary, and sufficient, to reproduce the experimental cross sections.
The reaction is therefore sensitive only to the ANC of the initial ground state and the phaseshifts in the $p$ continuum.
On carbon, the degrees of freedom corresponding to the $d$ continuum and the excitation of the core to its first $2^+$ state play a non-negligible role in the vicinity of the $\fial^+$ and $\thal^+$ states.
Second, because we have fitted the LECs of that NLO description upon \emph{ab initio} theoretical predictions, the very good agreement observed with experiment at both intermediate and relativistic beam energies, and on light and heavy targets, confirms the quality of these predictions.
In the next section, we explore how such Halo-EFT description of $^{11}$Be compares in transfer calculations.

\section{Transfer}\label{Trans}

\begin{figure}
\begin{minipage}{0.5\textwidth}
  \includegraphics[width=\textwidth]{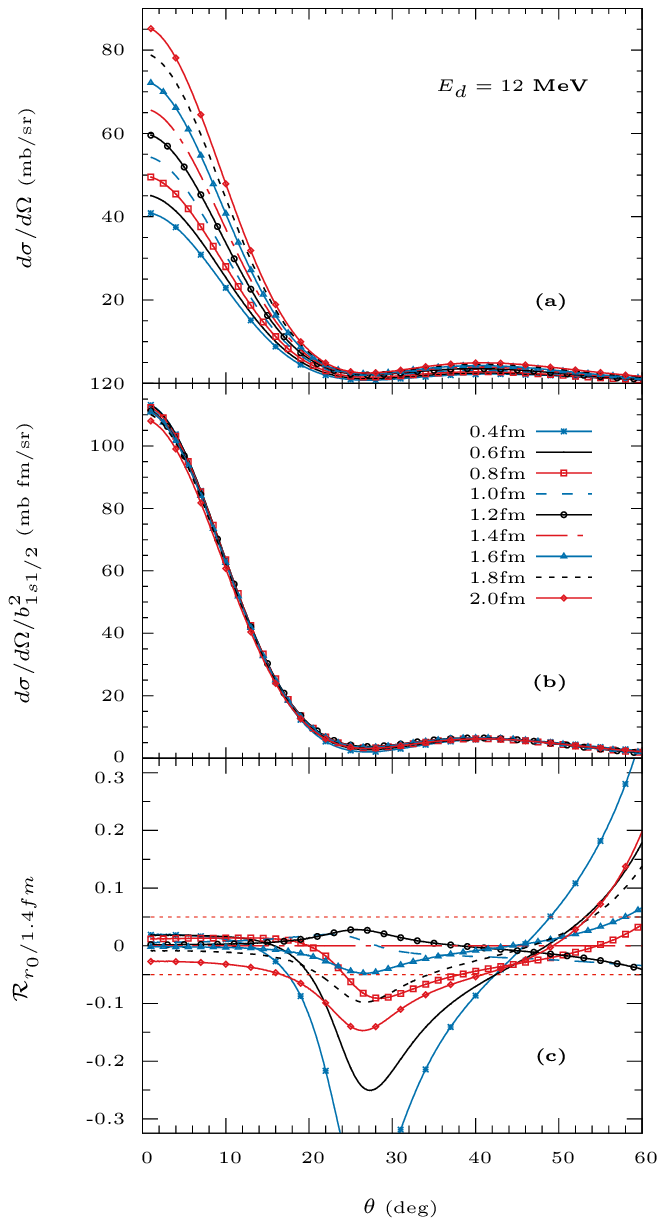}
  \includegraphics[width=\textwidth]{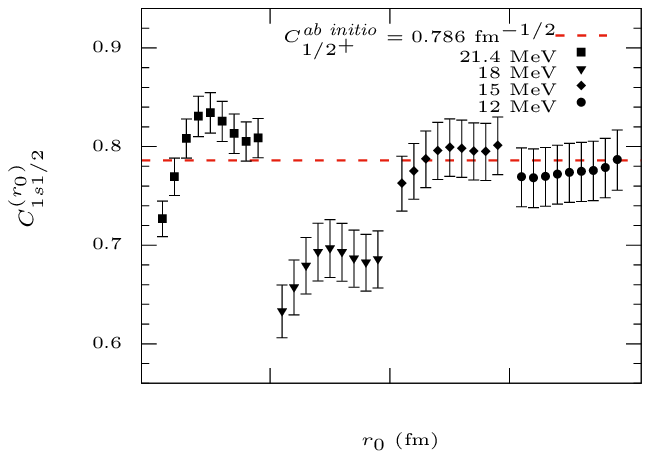}
\end{minipage}
\begin{minipage}{0.5\textwidth}
  \includegraphics[width=\textwidth]{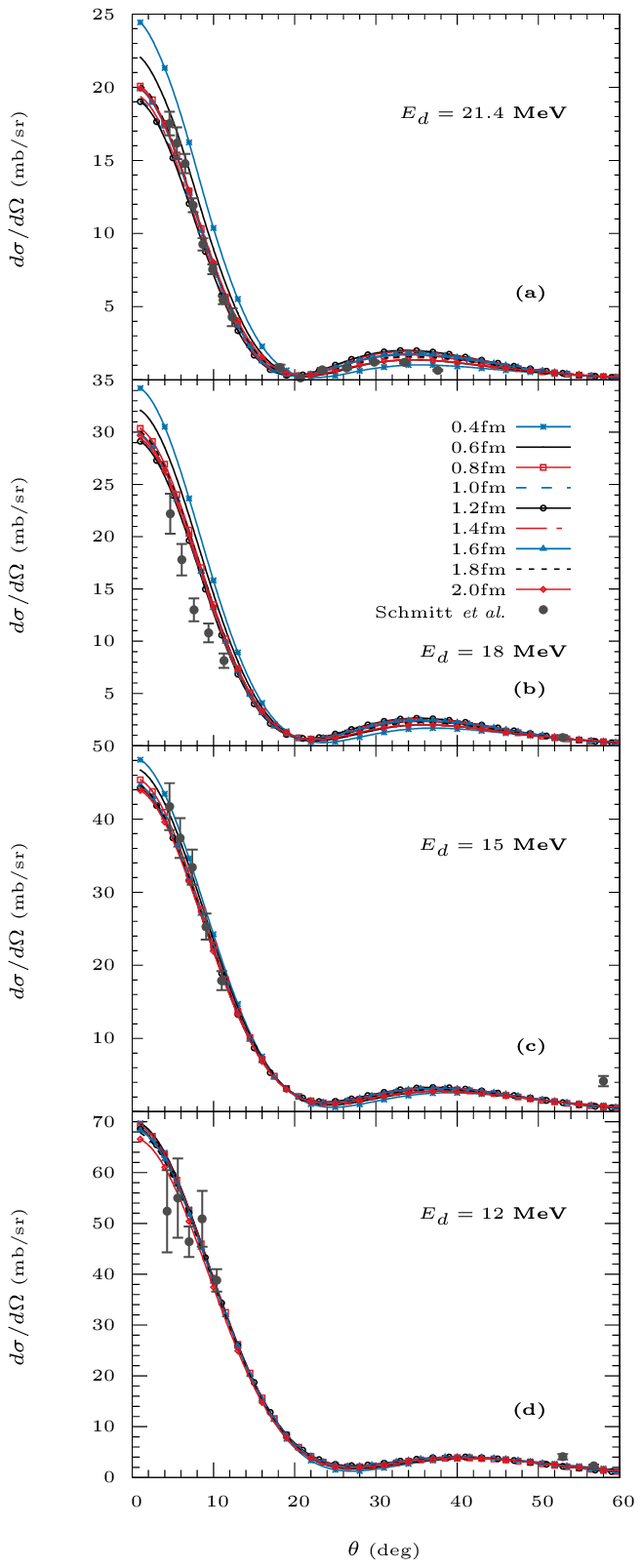}
\end{minipage}
\caption{Halo-EFT analysis of the $^{10}$Be(d,p)$^{11}$Be transfer reaction.
Top left panel (a): transfer cross section at an incoming deuterium energy $E_{\rm d}=12$~MeV
computed using LO $^{10}$Be-n effective interactions with ranges $r_0=0.4$--2.0~fm.
Left panel (b): same cross section normalised to the square of the single-particle ANC $b_{1s1/2}$ of the $^{11}$Be bound state.
Left panel (c): ratio ${\cal R}$ of the normalised cross sections, which shows that the reaction is purely peripheral up to about $20^\circ$. Within that angular range, an ANC for the $^{11}$Be ground state can be safely inferred.
Bottom left panel: the ANC deduced from the experimental data plotted as a function of $r_0$ for the four beam energies of Refs.~\cite{Sch12,Sch13}.
The dependence on $r_0$ is vanishing at low energy, showing that for $E_{\rm d}\le15$, the reaction is not sensitive to the short-range physics of the halo structure of $^{11}$Be, and hence that the corresponding ANCs are reliable.
They are in excellent agreement with the \emph{ab initio} prediction of Calci \etal\  \cite{CNR16}.
Right panels, the transfer calculations scaled to the inferred ANC are compared to the data of Schmitt \etal\ \cite{Sch12,Sch13}.
The agreement is excellent in the peripherality zone, so at forward angles and low energy $E_{\rm d}$.
Reprinted figures with permission from Ref.~\cite{YC18} Copyright (2018) by the American Physical Society.
}
\label{f4}
\end{figure}

The transfer $^{10}$Be$(\rm d,p)$$^{11}$Be has been measured in inverse kinematics at the Oak-Ridge National Laboratory with an ultra-pure $^{10}$Be beam at energies corresponding to direct-kinematics deuteron energies $E_{\rm d}=21.4$, 18, 15, and 12~MeV \cite{Sch12,Sch13}.
The measured angular distributions of the outgoing proton are displayed in the right panel of \Fig{f4}.

In this section, we summarise the work published in Ref.~\cite{YC18}, where we have reanalysed these data using a Halo-EFT description of $^{11}$Be.
To this aim, we resort to the usual three-body model of the reaction with an incoming deuteron d---seen as a neutron n bound to a proton p---impinging on a $^{10}$Be, whose internal structure is ignored \cite{GCM14}.
During the reaction, the neutron is transferred to the target to form a bound state of $^{11}$Be---seen as a neutron loosely bound to $^{10}$Be---while the proton is scattered away.
We choose the Adiabatic Distorted Wave Approximation (ADWA) in its finite-range implementation developed by Johnson and Tandy \cite{JT74,GCM14}.
This model of transfer includes the breakup of the deuteron at the adiabatic level.
It has been shown to provide a good description of such reactions \cite{UDF12}.

We have just seen that the breakup of $^{11}$Be is purely peripheral on both light and heavy targets, when it is performed at intermediate and relativistic beam energy.
Our goal in Ref.~\cite{YC18} was to determine how much  transfer reactions populating halo nuclei are peripheral.
To answer this question, a series of LO Halo-EFT descriptions has been computed.
As explained in \Sec{HaloEFT}, at this order, the $^{10}$Be-n interaction is described by a simple Gaussian term in the $s_{1/2}$ partial wave [see \Eq{e2}], whose depth is adjusted to reproduce the experimental binding energy.
Varying the range $r_0$ of that Gaussian, we obtain $c$-n radial wave functions with different ANCs, see \Fig{f1}(a); we consider $r_0$ between 0.4~fm and 2~fm with an incrementing step of 0.2~fm.
The corresponding ADWA cross sections are illustrated for the transfer to the $^{11}$Be ground state at $E_{\rm d}=12$~MeV in panel (a) of the left side of \Fig{f4}.
Once divided by the square of the single-particle ANC of the final state $b_{1s1/2}$, all the curves end up very close to each other (panel (b) of the left side of \Fig{f4}).

Because the cross section changes by orders of magnitude over the angular range considered in the experiment, we take the ratio of the cross sections divided by $b_{1s1/2}^2$ by that observable with $r_0=1.4$~fm, which is in the middle of the considered range of $r_0$.
These ratios are shown in the panel (c) of the left side of \Fig{f4}.
At forward angle, viz. up to $\theta\approx 20^\circ$, they differ by less than 5\% (horizontal red dashed lines).
At larger angles, the relative differences increase significantly.
From this we can deduce that when the data are selected at forward angles, the reaction is purely peripheral, and hence that an ANC for the final halo state can be safely inferred from transfer experiments. 
Our detailed analysis has shown that this ``peripherality region'' increases at low beam energy: the lower $E_{\rm d}$, the broader it becomes \cite{YC18}.
The ANCs deduced from the data in that peripheral region are shown in the bottom panel of the left side of \Fig{f4} for each beam energy.

Whereas the value of the ANC strongly depends on $r_0$ at $E_{\rm d}=21.4$ and 18~MeV\footnote{The ANCs inferred at $E_{\rm d}=18$~MeV are significantly lower than at other energies. At that energy, the calculations systematically overestimate the data (see the panel (b) of the right side of \Fig{f4}). This issue was noted in the original analysis of Schmitt \etal\ \cite{Sch12}, pointing to some systematic uncertainty. This issue has little effect on our study, since we have not considered this set because of the strong dependence of the ANC on $r_0$.}, it is nearly independent of the range of $V_{c\rm n}$ at $E_{\rm d}=15$ and 12~MeV.
At these low energies, this ANC~$=0.785\pm0.03$~fm$^{-1/2}$, which agrees nearly perfectly with the \emph{ab initio} prediction of Calci \etal\ \cite{CNR16} (ANC~$=0.786$~fm$^{-1/2}$; horizontal red dashed line).
This confirms what has been already mentioned in \Sec{BU} on breakup: these NCSMC calculations seem very reliable, since we recover nearly the same ANC from our independent analysis of transfer reactions.
A NLO description of $^{11}$Be would lead to a very good agreement with the data at forward angle and low beam energies, as illustrated by the plots in the right side of \Fig{f4}, where our ADWA calculations, scaled to the ANC inferred from the data, are compared to the experimental cross sections.

Note that here too the choice of optical potentials between the proton and neutron of the deuteron with $^{10}$Be affects the calculation.
However this change remains small.
Another choice of potential leads to ANC$=0.755\pm0.03$~fm$^{-1/2}$, which is still in very good agreement with the \emph{ab initio} prediction \cite{YC18}.

This Halo-EFT analysis of transfer reactions confirms the results of our analysis of the breakup of $^{11}$Be.
First, such a description enables us to easily deduce which structure observable affects the cross section, and hence, which information about the nucleus can be reliably inferred from the data.
Second, it can be used to test predictions of \emph{ab initio} structure calculations on reaction measurements.
In the next section, this analysis is extended to knockout.

\section{Knockout}\label{KO}

In knockout reactions (KO), the one-neutron halo projectile is sent at intermerdiate to high energy, viz. at or above $50A$~MeV, onto a light target, such as $^{9}$Be or $^{12}$C, and the events of interest are those in which the halo neutron is removed from the nucleus, viz. the remaining core is the only particle detected \cite{HT03}.
This inclusive measurement contains two contributions: the diffractive breakup, in which the neutron simply dissociates from the core, as seen in \Sec{BU}, and the stripping, in which the neutron is absorbed by the target.
Besides the total KO cross section, the observable that is usually measured is the momentum distribution of the core parallel to the beam axis.
This observable provides key information about the orbital angular momentum of the knocked-out neutron \cite{HT03,Aum00}.
Most often, a spectroscopic factor for the dominant $c$-n configuration in the projectile structure is also deduced from the data.

\begin{figure}
\center
  \includegraphics[width=0.5\textwidth]{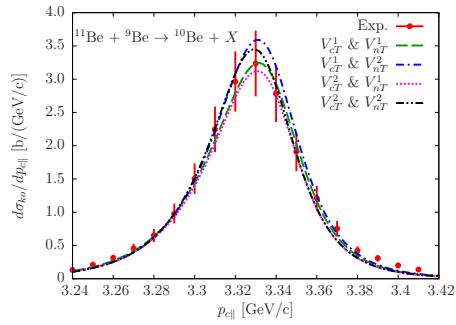}
\caption{One-neutron removal cross section for $^{11}$Be on a Be target at $60A$~MeV.
The cross section is expressed as a function of the $^{10}$Be momenta parallel to the beam axis.
Eikonal calculations performed with the NLO description of the projectiles developed in Ref.~\cite{CPH18} (with range $r_0=1.2$~fm) are compared to the data of Ref.~\cite{Aum00}.
Results with different choices of optical potentials show the sensitivity to these interactions, which clearly exceeds the tiny 1.5~\% difference due to $r_0$ \cite{HC21}.
Reprinted figures with permission from Ref.~\cite{HC21} Copyright (2018) by the American Physical Society.
}
\label{f5}
\end{figure}

In the present section, we turn to Halo-EFT to reanalyse the data gathered at MSU for $^{11}$Be impinging on $^{9}$Be at $60A$~MeV by Aumann \etal\ \cite{Aum00}, see \Fig{f5}.
As in \Sec{BU}, we compute the diffractive breakup part of the cross section with the DEA \cite{BCG05,GBC06}.
The stripping contribution is estimated at the usual eikonal approximation, i.e., considering the adiabatic approximation \cite{HT03}.
As in the previous sections, we consider a Halo-EFT description of $^{11}$Be, we use the same NLO $^{10}$Be-n potentials as in the breakup analyses \cite{CPH18,MC19}.

In a first, purely theoretical, study, we have observed that KO observables for one-neutron halo nuclei are also purely peripheral, in the sense that they are insensitive to the short-range physics of the projectile structure \cite{HC19}.
In particular, they depend only on the ANC of the initial bound state and not on its norm, i.e., they are mostly independent of the spectroscopic factor.
In the other $c$-n partial waves, the KO cross section slightly depends on the existence of excited bound states.
The KO cross section is reduced when other bound states are included in the projectile description.
This indicates that a proper Halo-EFT description of $^{11}$Be in KO models should go at least up to NLO to account for its $\half^-$ excited state.
Interestingly, the structure of the continuum, in particular the presence of single-particle resonances, does not affect the KO cross section \cite{HC19}.
Therefore, as for the Coulomb breakup, the NLO description of $^{11}$Be should also be sufficient to describe this reaction.
Because the $p$ continuum has little influence on the cross section, this reaction is well suited to infer the ANC of the ground state, as done, e.g., in Ref.~\cite{TAM02}.

Following this preparatory work, we have reanalysed the MSU KO data with a NLO Halo-EFT description of $^{11}$Be \cite{HC21}.
The results of our calculations are illustrated in \Fig{f5} alongside the measurements of Ref.~\cite{Aum00}.
The NLO description of $^{11}$Be chosen in this case corresponds to the Gaussian range $r_0=1.2$~fm (see \Sec{HaloEFT}).
The sensitivity of the total KO cross section to that choice is lower than 1.5\%, hence negligible compared to the 15\% experimental uncertainty \cite{Aum00}.
To estimate the sensitivity of our calculations to the $c$-$T$ and n-$T$ interactions, we use different optical potentials found in the literature \cite{HC21}.

We observe an excellent agreement with experiment: all calculations fall within the uncertainty range of the data.
Interestingly, our calculations reproduce most of the asymmetry of the data, which is a sign that our DEA calculation of the diffractive breakup properly accounts for the dynamics of the reaction \cite{CBS08,HC21}.
We note the large sensitivity of this KO observable to the optical potentials, specially to $V_{{\rm n}T}$.
Though smaller than the experimental uncertainty, this sensitivity clearly exceeds that related to the range $r_0$ of the $V_{c\rm n}$ potential.
A precise determination of the projectile ANC hence requires a significant improvement of these interactions.
From our calculations, we can infer an ANC~$=0.79\pm0.04\pm0.06$~fm$^{-1/2}$, 
where the first uncertainty corresponds to the choice of optical potentials and the second translates the experimental uncertainty.
This is, once again, in excellent agreement with the prediction of Calci \etal\  of ANC~$=0.786$~fm$^{-1/2}$ \cite{CNR16}.
This analysis constitutes the last piece of the puzzle and completes our systematic study of reactions involving $^{11}$Be using Halo-EFT to describe that archetypical one-neutron halo nucleus.
This confirms the interest of using Halo-EFT in modelling reactions with halo nuclei \cite{HC19}.
Moreover, the excellent agreement obtained in each and every case confirms the quality of the \emph{ab initio} calculations of Calci \etal\ \cite{CNR16,HC21}.

\clearpage

\section{Conclusion}\label{Conclusion}

In this short review, I have illustrated the interest of using Halo-EFT to describe one-neutron halo nuclei within precise models of reactions.
The first advantage is that through its expansion, Halo-EFT provides a natural order of importance of the structure observables \cite{BHvK02,BHvK03,HJP17}.
Our tests have shown that this order is also observed at the level of reaction observables.
Halo-EFT therefore helps us identify the nuclear-structure observable that matter in the description of reactions, and, accordingly, that are probed by these reactions.
Weinberg's seminal idea \cite{Wei90,Wei91} is thus also confirmed in the description of reactions involving halo nuclei, for which a clear separation of scales is observed.

The main conclusion of the analyses performed on breakup \cite{CPH18,MC19}, transfer \cite{YC18}, and knockout \cite{HC19,HC21} is that these reactions are mostly peripheral, in the sense that they probe only the tail of the projectile wave function, and not its interior.
The structure observables that matter in the examples shown here on $^{11}$Be are the binding energy of the halo neutron to the core, the ANC of the nucleus bound states, and, in breakup, the phaseshift in the $p$ continuum.
This confirms anterior results on breakup reactions \cite{CN06,CN07}.
Accordingly, a NLO description of $^{11}$Be is therefore necessary, and sufficient, to compute the cross sections of most of the reactions involving $^{11}$Be.
Note that this conclusion holds also for $^{15}$C, another well-known one-neutron halo nucleus \cite{MYC19,HC21}.

In addition, by adjusting the LECs of the EFT expansion upon \emph{ab initio} predictions, we can directly confront the quality of these calculations to reaction observables.
On the particular case of $^{11}$Be, since we have observed excellent agreements between our calculations and so many different experimental data using NLO descriptions of $^{11}$Be constrained by the \emph{ab initio} calculation of Calci \etal\ \cite{CNR16}, we can safely conclude to the quality of these calculations.
Similar conclusions can be drawn for $^{15}$C \cite{MYC19,HC21}.

In some cases, in particular in nuclear-dominated reactions, the calculations are marred with the uncertainty related to the optical potentials chosen to simulate the interaction between the projectile constituents and the target.
Improving the reliability of these potentials would help us better analyse experiments.
Interestingly, there too EFT can provide some help \cite{Hol13,Vor16,Rot17,Dur18,Idi19,DCS20}.
However, as Kipling said, that's another story\ldots

In conclusion, combining a Halo-EFT description of nuclei with precise nuclear-reaction models is an excellent way to investigate the structure of halo nuclei through reactions and confront nuclear-structure calculations with experiment.
In the future, we plan to pursue this avenue and extend this idea to other exotic nuclei, such as $^{31}$Ne.

\begin{acknowledgements}
This project has received funding from the European Union’s Horizon 2020 research	and innovation program under grant agreement No 654002, the Deutsche Forschungsgemeinschaft Projekt-ID 279384907 – SFB 1245 and Projekt-ID 204404729 – SFB 1044, and the PRISMA+ (Precision Physics, Fundamental Interactions and Structure of Matter) Cluster of Excellence.
I acknowledge the support of the State of Rhineland-Palatinate.
\end{acknowledgements}


\end{document}